\documentclass{article}
\pdfoutput=1

\usepackage{arxiv}

\usepackage[utf8]{inputenc} 
\usepackage[T1]{fontenc}    
\usepackage{hyperref}       
\usepackage{url}            
\usepackage{booktabs}       
\usepackage{amsfonts}       
\usepackage{nicefrac}       
\usepackage{microtype}      
\usepackage{graphicx}
\usepackage[caption=false]{subfig}
\usepackage{doi}
\usepackage{multirow}
\usepackage{multicol}

\title{Automatic Metadata Extraction Incorporating Visual Features from Scanned Electronic Theses and Dissertations}

\author{Muntabir Hasan Choudhury\\
	Old Dominion University\\
	Norfolk, VA 23529 \\
	\texttt{mchou001@odu.edu} \\
	\And
    Himarsha R. Jayanetti\\
    Old Dominion University\\
	Norfolk, VA 23529 \\
	\texttt{hjaya002@odu.edu} \\
	\And
    Jian Wu\\
	Old Dominion University\\
	Norfolk, VA 23529 \\
	\texttt{j1wu@odu.edu} \\
	\And
	William A. Ingram \\
	Virginia Polytechnic Institute and State University \\
	Blacksburg, VA 24061 \\
	\texttt{waingram@vt.edu} \\
	\And
	Edward A. Fox \\
	Virginia Polytechnic Institute and State University \\
	Blacksburg, VA 24061 \\
	\texttt{fox@vt.edu} \\
}

\renewcommand{\shorttitle}{Automatic Metadata Extraction Incorporating Visual Features from Scanned ETDs}

\begin{document}
\maketitle

\begin{abstract}
Electronic Theses and Dissertations (ETDs) contain domain knowledge that can be used for many digital library tasks, such as analyzing citation networks and predicting research trends. Automatic metadata extraction is important to build scalable digital library search engines. Most existing methods are designed for born-digital documents, so they often fail to extract metadata from scanned documents such as for ETDs. Traditional sequence tagging methods mainly rely on text-based features. In this paper, we propose a conditional random field (CRF) model that combines text-based and visual features. To verify the robustness of our model, we extended an existing corpus and created a new ground truth corpus consisting of 500 ETD cover pages with human validated metadata. Our experiments show that CRF with visual features outperformed both a heuristic and a CRF model with only text-based features. The proposed model achieved 81.3\%-96\% F1 measure on seven metadata fields. The data and source code are publicly available on Google Drive\footnote{https://tinyurl.com/y8kxzwrp} and a GitHub repository\footnote{https://github.com/lamps-lab/ETDMiner/tree/master/etd\_crf}, respectively.
\end{abstract}


\section{Introduction}
A thesis or dissertation is one type of scholarly work that shows a student pursuing higher education has successfully met key requirements of a degree. An ETD is usually accessible from a university's digital library or a third-party ETD repository such as ProQuest. Since the inception of ETDs around 1997, pioneered by Virginia Tech, many ETDs are generated electronically (i.e., born-digital) by computer software such as LaTeX and Microsoft Word. However, the majority of the ETDs produced before 1997 and a significant fraction of ETDs after 1997 are scanned from physical copies (i.e., non-born digital). These ETDs are valuable for digital preservation, but to make them accessible, it is necessary to index the metadata of these ETDs. 

Many ETD repositories are accompanied by incomplete, little, or no metadata, posing challenges for accessibility. For example, advisor names appearing on the scanned ETDs may not be available in the metadata provided in the library repository. Thus, an automatic approach should be considered to extract metadata from scanned ETDs. Many tools \cite{Lipinski-2013,han-2003,lopez2009grobid,cermine} have been developed to automatically extract metadata for relatively short and born-digital documents, such as conference proceedings and journals published in recent years. However, they do not work well with scanned book-length documents such as ETDs. Extracting metadata from scanned ETDs is challenging due to poor image resolution, typewritten text, and imperfections of the OCR technology. Many commercial-based OCR tools such as OmniPage, ABBYY FineReader, or Google Cloud API OCR could be used for converting PDFs to text, but they usually incur a cost.  Therefore, we adopted Tesseract-OCR, an open-source OCR tool, to extract metadata from the cover pages of scanned ETDs. Tesseract-OCR supports printed and scanned documents and more than 100 languages.
It returns output in plain text, hOCR, PDF, and other formats.

In our preliminary work \cite{choudhury-jcdl2020}, we proposed a heuristic method to extract metadata from the cover pages of scanned ETDs. However, heuristic methods usually do not generalize well. They often fail when applied to a set of data with a different feature distribution. In this paper, we investigate the possibility of improving the generalizability of our method based on a learning-based model.

\section{Related Work}
Many frameworks have been proposed to extract metadata from scholarly papers. CERMINE \cite{cermine} was developed to extract structured bibliographic data from scientific articles. It can extract information related to title, author, author's affiliation, abstract, keywords, journal, volume, issue, pages, and year. For the metadata extraction tasks, they used both machine learning models such as Support Vector Machine (SVM) and simple rule based models. CERMINE achieved an average F1 score of 77.5\% for most metadata types and the benchmark evaluation outperformed the existing tools, including GROBID \cite{lopez2009grobid} and ParCit \cite{councill2008parscit}, while extracting metadata such as title, email addresses, year, and references. One limitation of this tool is that the PDF documents which contain scanned pages in the form of images will not be properly processed.

GROBID \cite{lopez2009grobid} is a text mining library for extracting bibliographic metadata from born-digital papers. GROBID is based on eleven different CRF models and each uses the same CRF-based framework which utilizes position (e.g., beginning or ending of the line), lexical information, and layout information. It is capable of extracting header and bibliographic metadata such as title, authors, affiliations, abstract, date, keywords, and references. It achieves an accuracy of 74.9\% per complete header instance but it fails to extract metadata from non-born digital documents such as the cover page of scanned ETDs.

In our previous work \cite{choudhury-jcdl2020}, we have introduced a heuristic model to extract metadata fields from scanned ETD cover pages. It is a rule based method where the metadata fields are captured using a set of carefully designed regular expressions. Table \ref{tab:rules2} shows the accuracy values obtained for each field for the sample of 100 ETDs.
These range from 39\% to 97\%.

\section{Dataset}
\begin{figure}[t]
\centering
   \includegraphics[width=0.8\textwidth]{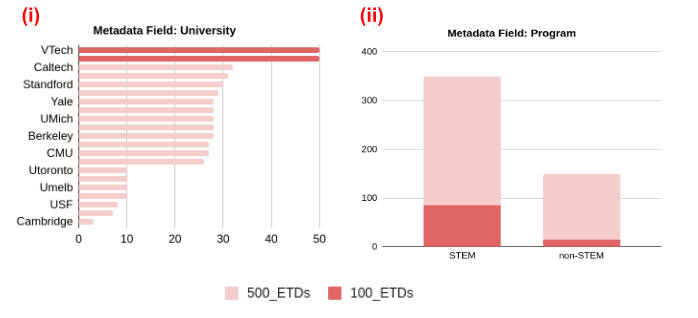}%
   \vspace{-7mm}
   \caption{Distribution of metadata fields: University (i) and Program (ii) in the corpus of 500 ETDs.}
   \label{fig:dataset} 
\end{figure}

The dataset used in our previous study \cite{choudhury-jcdl2020} consisted of a relatively small number of ETDs from only two universities. To overcome this limitation, we created a new dataset of 500 ETDs, which includes 450 ETDs from 15 US universities and 50 ETDs from 6 non-US universities as illustrated in Figure~\ref{fig:dataset}. These ETDs were published between 1945 and 1990. There are 350 STEM and 150 non-STEM majors from 468 doctoral, 27 master's, and 5 bachelor's degrees. We derived the following seven intermediate datasets from our set of 500 ETDs.

\begin{enumerate}
    \item The cover page of each ETD in PDF format. 
    \item TIFF images of (1). The TIFF format is used as the input to Tesseract because it tends to produce fewer errors than JPEG.
    \item TXT-OCR: The output of Tesseract containing noisy text extracted from the TIFF images.
    \item TXT-clean: The cleaned version of the TXT-OCR dataset after correcting misspellings, fixing the mistakes during OCR, lowercasing the text, and removing empty lines between text. We did not remove line breaks.
    \item TXT-annotated: Seven metadata fields annotated using the GATE annotation tool \cite{gate}.
    \item GT-meta: The ground truth from metadata provided by libraries. The data were gathered in the XML-format from MIT, JSON-format from Virginia Tech, and in HTML format for all other universities from the ProQuest database.
    \item GT-rev: Revised metadata from GT-meta after manually rectifying discrepancies between library provided metadata and the data present in the cover page of PDF documents. 
\end{enumerate}

\begin{figure}[t]
    \centering
   \includegraphics[width=0.6\textwidth]{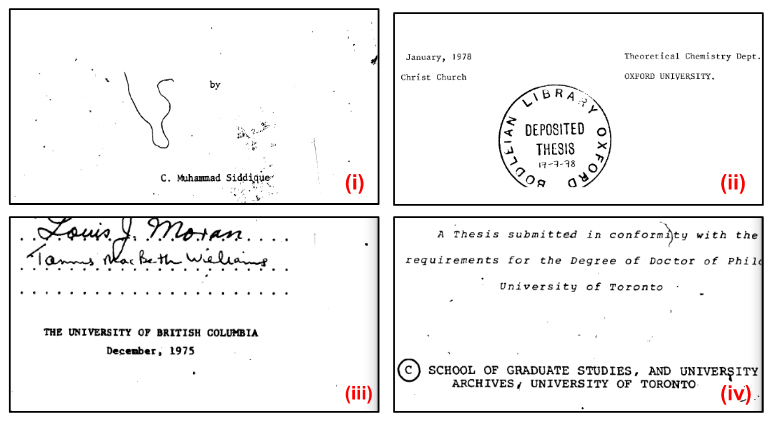}%
   \caption{OCR challenges for ETDs: scribble (i), stamp (ii), overlapped letters (iii), and copyright character (iv).}
   \label{fig:challenges}
\end{figure}

We observed several challenges to convert scanned ETDs to text (Figure~\ref{fig:challenges}). 

\begin{enumerate}
    \item Some fields were not available on the cover page.
    \item Lines were present to fill the title, degree, author, etc.  
    \item Multiple years are provided, such as ``submitted year'' and ``publication year.''
    \item There were ETD cover pages where author's previous educational certifications are listed (e.g., University of British Columbia) making it difficult to extract the degree field. 
    \item College name is mentioned instead of university name (e.g., University of Oxford).
\end{enumerate}

\section{Methodology}
\subsection{CRF with Sequence Labeling (CRF Model)}
CRF is a statistical modeling algorithm. This model assumes that the features are dependent on each other, but also considers future observation when modeling a sequence. It encodes each token of the annotated fields as the beginning (B) and inside (I). For example, if the token represents an author's name, we will tag it as B-author and I-author. The tokens which are not a part of the metadata fields should be tagged as outside (O). The BIO tagging schema has been applied in studies such as named entity recognition \cite{hesdk-jianwu}\cite{councill2008parscit} and keyphrase extraction \cite{sujatha-keyphrase}. 

We tagged each token with Part of Speech (POS). POS tags are important here if the phrase consists of pronoun, preposition, or determiner, e.g., ``at the Massachusetts Institute of Technology.'' If the current token is ``Massachusetts'' tagged as NNP, we can infer the POS of the two tokens before the current token. This assigns to the previous two tokens, ``at'' and ``the,'' the POS tags IN and DT, respectively. Our model extracts the following features.

\begin{enumerate}
    \item Whether all the characters in the string are uppercase.
    \item Whether all the characters in the string are lowercase.
    \item Whether all the characters in a string are numeric. 
    \item Last three characters of the current word.
    \item Last two characters of the current word.
    \item POS tag of the current word.
    \item Last two characters in the POS tag of the current word.
    \item POS of the two tokens after the current word. 
    \item Whether the first character of consecutive words is uppercase. For example, a title would have consecutive words that start with an uppercase letter.
    \item Whether the first character is uppercase for the word that is not at the beginning or end of the document. This is useful for metadata fields such as author, advisor, program, degree, and university. These fields are not generally at the beginning or end of the document.
\end{enumerate}

\subsection{CRF with Visual Features (CRF-visual)}
In the heuristic and the CRF models, we only incorporate text-based features. When humans annotate the documents, they not only rely on text, but also visual features, such as the positions of the text and their lengths. For example, thesis titles usually appear in the upper half of the cover page and the authors and advisors usually appear in the lower half of the cover page. This inspires us to investigate whether incorporating visual features can improve performance.

\begin{figure}[htp]
    \centering
    \subfloat[\centering]{{\includegraphics[width=10.5cm]{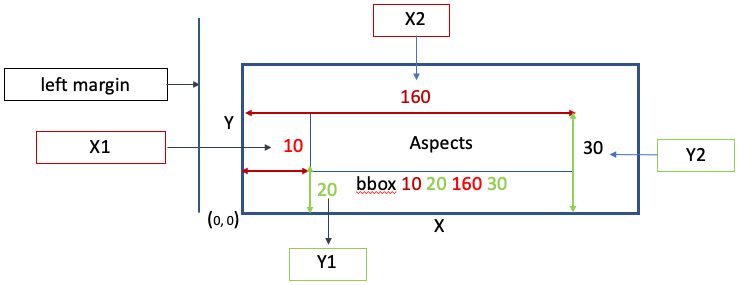} }}
    \newline
    \subfloat[\centering]{{\includegraphics[width=10.5cm]{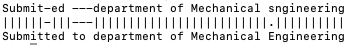} }}
    \caption{(a) Bounding box measurement (b) OCR output text (i.e., noisy) alignment with clean text}%
    \label{fig:bbox and text-alignment}%
\end{figure}

Visual information is represented by corner coordinates of the bounding box (bbox) of a text span. We extract all x-coordinate values (e.g., x1, x2) and y-coordinate values (e.g, y1, y2) for each token. This information is available from hOCR files and XML files, which are output from Tesseract. Figure~\ref{fig:bbox and text-alignment}(a) illustrates the bounding box information of the token with x and y coordinates. x1 is the distance from the left margin to the bottom right corner of bbox, y1 is the distance from the bottom margin to the bottom right corner of bbox, x2 is the distance from the left margin to the upper right corner of bbox, and y2 is the distance from the bottom margin to the upper right corner of the bbox. All coordinates are measured with reference to the bottom-left corner of the token.

However, transferring these visual features is challenging because the ground truth text is output by Tesseract and rectified by humans. Therefore, the characters in the rectified text are not necessarily aligned with Tesseract's output. The position information was only available for the OCR output. We applied text alignment using the longest common sequence \cite{sequence-align}. In Bioinformatics, sequence alignment has been commonly applied to align protein, DNA, and RNA sequences which are usually represented by a string of characters. We used an open-source tool known as Edlib \cite{edlib} to align the noisy text data and clean text. Edlib computes the similarity and minimum edit distance between two text sequences. Then we map the positions for each token from TXT-OCR to TXT-clean. Figure~\ref{fig:bbox and text-alignment}(b) illustrates an example of the alignment. We incorporated three visual features to enhance the performance of the CRF model. The following three features have been normalized between 0 and 1.

\begin{enumerate}
    \item left margin --- x1 as a feature for all tokens in the same line.
    \item upper left corner --- y2 as a feature for all tokens.
    \item bottom right corner ---  y1 as a feature for all tokens. 
\end{enumerate}

\subsection{BiLSTM-CRF Model}
Bidirectional Long Short Term Memory (BiLSTM) can learn the hidden features automatically.
It has been proven to be effective in sequence labeling problems \cite{dke-wu_reshad} and in encoding sequence tokens into fixed-length vectors. BiLSTM tries to learn the context of the given sentence in both forward and backward directions. We added a CRF layer that classifies tokens based on their encoding. We investigated BiLSTM-CRF to extract seven metadata fields. The architecture of the classifier consists of three components: a word-embedding layer, a BiLSTM as an encoder, and a CRF layer. BiLSTM learns the forward and backward context in a sequence and feeds it into the CRF layer, which further predicts the labels for each token in a sequence. We used Adam as the optimizer and Keras word embedding initialized with random weights. The batch size is set to 32, and the model runs for up to ten epochs. 

\begin{table}
    \centering\small 
    \caption{Performance (accuracy) comparison between the heuristic model on two datasets.}
    \begin{tabular}{c|c|c}
    \toprule
      \textbf{Field} & {\bf Accuracy\% (100)} & {\bf Accuracy\% (500)} \\
     \midrule
     {Title} & {81\%}  & {45.0\%} \\
     \midrule
      {Author} & {78\%} & {62.8\%} \\
     \midrule
      {Degree} & {81\%} & {58.0\%}\\
     \midrule
      {Program} & {97\%} & {8.0\%} \\
     \midrule
      {Institution} & {94\%} & {18.8\%} \\
     \midrule
      {Year} & {65\%} & {37.8\%} \\
     \midrule
     {Advisor} & {36\%} & {5.0\%} \\
     \bottomrule
    \end{tabular}
    \label{tab:rules2}
\end{table}

\section{Evaluation and Results}
\subsection{Heuristic Model}
We apply the heuristic method \cite{choudhury-jcdl2020} on 500 ETDs (Table~\ref{tab:rules2}). The accuracy of the heuristic method on the 500 ETDs is considerably lower than the accuracy in the 100 ETDs in the previous study \cite{choudhury-jcdl2020}. This is because the new dataset contains ETDs from a more diverse set of universities and majors.

\subsection{CRF Model and CRF-Visual}
\begin{figure}[t]
    \centering
    \includegraphics[width=10cm]{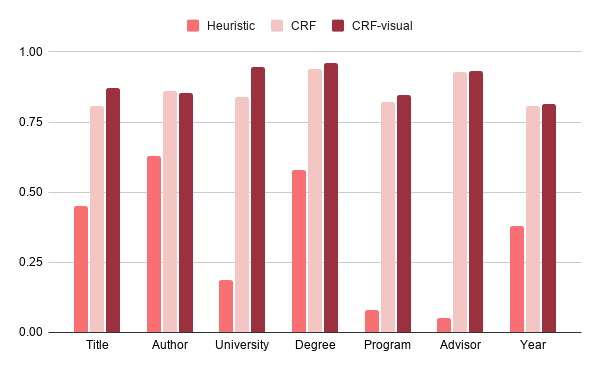}
    \caption{Performance (F1) comparison of the models}
    \label{fig: three_models}
\end{figure}

We randomly divide the samples into two sets: 350 samples for training and 150 samples for testing. The CRF model predicts the labels for each metadata field at the token level. However, we must glue together the predicted tokens for each metadata field and compare them against the corresponding GT-meta and GT-rev. While comparing the title field against the GT-meta and GT-rev, a fraction of predicted titles did not match exactly, with a small difference such as a punctuation mark or a space character. For example, the model predicted the title as ``\underline{thermo- fluid} dynamics of separated \underline{two - phase} flow.'' However, in the GT meta it is ``\underline{thermo-fluid} dynamics of separated \underline{two-phase} flow.'' These small offsets are not caused by the model but by line breaks and additional punctuation marks added in text justification. Therefore, the predicted span should be counted as a true positive. We use a fuzzy matching algorithm based on Levenshtein distance and set a threshold of 0.95 when matching predicted and ground truth titles. Figure~\ref{fig: three_models} illustrates the performance of our model. The model outperformed the baseline model whereas CRF-visual outperformed both the baseline model and CRF. 

\subsection{BiLSTM-CRF Model}
The BiLSTM-CRF model generated poor results for all fields. The F1 scores for token level labels such as B-title, I-title, B-author, and I-author were only 34\%, 34\%, 24\%, and 23\%, respectively. The F1 measures for the remaining fields were even lower, so we did not plot them in Figure~\ref{fig: three_models}. There are several possible reasons. One major reason is the small size of the training data. The training set contains 350 ETD cover pages. Some fields contain less than 100 samples. This is likely to overfit the neural model, so it does not generalize well. Another reason could be due to the default word embedding model provided by Keras. In light of recent advances in pre-trained language models that rely on contextualized word embeddings \cite{bert}, it is possible to fine-tune these models on a relatively small set of training data, which is a promising approach to beat the CRF model. 

\section{Conclusion}
We applied three models including CRF, CRF-Visual, and BiLSTM-CRF to extract metadata appearing on the cover pages of ETDs. We found that incorporating visual features into the CRF model boosts the F1 by 0.7\% to 10.6\% for seven metadata fields. The BiLSTM-CRF model did not perform well, which was likely due to insufficient training data and/or a lack of semantic features. We will enhance this model by adding pre-trained language models such as BERT \cite{bert}. In the future, we will also add post-OCR error correction into our pipeline and run directly on real data (i.e., TXT-OCR) instead of rectified data (i.e., TXT-clean). 

\section*{Acknowledgement}
Support was provided by the Institute of Museum and Library Services through grant LG-37-19-0078-198.

\bibliographystyle{unsrt}
\bibliography{references}  






\end{document}